\documentclass[useAMS,usenatbib,usegraphicx]{mn2e}
\usepackage{bm}
\usepackage{times}

\title[Far-infrared mapping of M\,81]
{Importance of far-infrared mapping in a spiral galaxy:
\textit{AKARI} observation of M\,81}
\author[Sun \& Hirashita]{Ai-Lei Sun$^{1,2}$
and Hiroyuki Hirashita$^1$\thanks{E-mail:
    hirashita@asiaa.sinica.edu.tw}
\\
$^1$Institute of Astronomy and Astrophysics, Academia Sinica,
P.O. Box 23-141, Taipei 10617, Taiwan\\
$^2$Department of Physics, National Taiwan University,
Taipei 10617, Taiwan
}
\date{2010 September 23}

\pagerange{\pageref{firstpage}--\pageref{lastpage}} \pubyear{2010}

\begin{document}
\label{firstpage}
\maketitle

\begin{abstract}
The importance of the far-infrared (FIR) mapping is
demonstrated for a face-on spiral galaxy, M\,81, by
analyzing its imaging data at 65, 90, and 140 $\micron$
taken by \textit{AKARI}. Basic products are the dust
temperature map, the dust optical depth map, and the
colour--colour diagram. The main features are as
follows. (i) The dust temperature derived from the
total fluxes at 90\,$\micron$ and 140\,$\micron$
reflects the relatively low temperatures seen in the
interarm and spiral arms excluding the warm spots,
rather than the high temperatures in warm spots and
the centre. This indicates that the total FIR luminosity
is dominated by the dust heated by the general
interstellar radiation field. 
(ii) The galaxy is more extended at 140 $\micron$
than at the other shorter wavelengths, which
reflects the radial dust temperature gradient.
(iii) The dust optical depth derived from the FIR mapping
is broadly consistent with that estimated from the
FIR-to-ultraviolet luminosity ratio.
(iv) The FIR colour--colour diagram is
useful to identify a `contamination' of warm dust.
The existence of
small-scale warm star-forming regions is supported
in the bright spots along the spiral
arms. This contamination also leads to an underestimate
of dust optical depth (or dust column density).
\end{abstract}


\section{Introduction}

The far-infrared (FIR) emission from galaxies is
dominated by the thermal emission from dust grains
\citep[e.g.][]{harper73}, and is known to be a good
indicator of star formation rate
\citep[e.g.][]{kennicutt98,inoue00}. Since the major
heating source of dust grains is usually stellar
radiation, the FIR colour, which reflects the dust
temperature, is one of
the most direct indicators of the interstellar
radiation field (ISRF). Moreover, the FIR emission is
usually optically thin in the ISM, which means that we
can trace all the radiation over the entire column.

{
Spatially resolved images of galaxies are required to
study the dust properties in different regions of a
galaxy and the characteristics of individual
star-forming regions. Also, since only integrated global
properties are observable for distant star-forming
galaxies, it is
important to clarify the relation between the global and
local characteristics. Specifically we should distinguish
whether the major dust heating comes from individual
star-forming regions or general ISRF}\footnote{The
general ISRF is 
defined as the average radiation field one
would measure from a location somewhere in a galactic
disc where the radiation is not dominated by any
particular star or star cluster \citep{walterbos96}.}
\citep[e.g.][]{walterbos96} 
{to clarify the major origin of the dust heating
source responsible for the FIR emission.
However, because the spatial resolution of FIR
observation is generally poor, only galaxies with large
angular size can be well resolved.}

M\,81 is particularly suitable for our purpose because
(i) it is a nearby big galaxy with a face-on geometry;
(ii) it can be compared with another well-known 
spiral galaxy, the Milky Way; and (iii) it is within
a field of view of the scan observation of
\textit{AKARI}, which is useful to quantify the
global properties and to subtract the
background.\footnote{In this paper, we use the
\textit{AKARI} data. Within the \textit{AKARI} archives,
M\,101 also satisfies our
requirements, but its FIS data are taken as part of the FIS
calibration program \citep{suzuki07} and are not open to
public (i.e., a special treatment for the data is
necessary). Other spiral
galaxies observed by \textit{AKARI} are too near or too far
for our aim.}

Which wavelengths in FIR are necessary? At long
wavelengths ($\ga 90~\micron$), the dust emission is
mainly contributed by relatively large grains in radiative
equilibrium with the stellar radiation field.
However, an emission excess in shorter wavelengths
($\la$ 65$\micron$) is usually observed, which
comes from very small dust grains suffering from
stochastic heating, that is, heated transiently to
a high temperature \citep{aannestad79,draine85}.
In order to estimate the large grain temperature,
which can be used to indicate the intensity of the ambient stellar
radiation, at least two bands are needed at wavelengths
$\ga 90~\micron$.
Furthermore, we require {three wavelengths} in the
FIR to {make} a FIR colour--colour {diagram}.
 {It has been shown that
the FIR colour--colour
relation of galaxies is useful to derive the
wavelength dependence of dust emissivity
\citep{dale01,nagata02,hibi06}.
The FIR colour--colour relations in the Galactic
disc and the Large/Small Magellanic Clouds 
are tight, indicating that the
wavelength dependence of dust FIR emission can be
considered to be uniform within and among
those galaxies \citep{shibai99,hibi06}. Such a
common wavelength dependence of dust emission
can be nicely reproduced by a common grain
optical properties and size distribution
\citep*{hirashita07}.
In this paper, the FIR colour-colour diagram also 
proves to be important for investigating
the dust emission characteristics.}

{
Slow-scan data of Far Infrared Surveyor (FIS) onboard
\textit{AKARI} \citep{murakami07,kawada07} with the 65,
90, and 140 $\micron$ bands is enough for the above
requirements. (We do not use the 160 $\micron$ data
because the data quality is not good enough for our
purpose.) The 90 and 140 $\micron$ bands are used to
estimate the large grain equilibrium temperature, and
a colour--colour diagram can be made with the 65, 90,
and 140 $\micron$ bands.
Besides, by using the \textit{AKARI} FIS data, a direct
comparison among M\,81 and a large number of galaxies in
the \textit{AKARI}
All-Sky data \citep{yamamura10} can be made because of
the common observational facility.
Very recently,\footnote{After the submission of this paper.}
\citet{suzuki10} has reported an analysis of the
\textit{AKARI} data of
M\,81. They focus on the
star formation law but do not investigate the details of the
dust temperature distribution within the galaxy and
the FIR colour--colour diagram.
M\,81 is also observed by the \textit{Spitzer Space
Telescope} \citep{perez06}, whose 70 $\micron$ and
160 $\micron$ data can be compared with our data.
However,
the dust temperature from
the 70 $\micron$ and 160 $\micron$ bands can
overestimate the
equilibrium temperature because of the possible
contamination in the 70 $\micron$ emission by the
stochastically heated very
small grains. Thus,
another FIR band at
$\ga 90~\micron$ is desirable.
This galaxy has recently been observed by
\citet{bendo10} with
\textit{Herschel Space Observatory} \citep{pilbratt10},
}
whose long-wavelength coverage is particularly useful to
examine the robustness of our analysis against the
inclusion of long-wavelength data.
{Although the \textit{Herschel} data are
better than our \textit{AKARI} data both in the spatial
resolution and sensitivity, we concentrate on the
\textit{AKARI} data in this paper because not only the
data of M\,81 is available but also the All-Sky Survey
data can be used to test if the FIR colours obtained for
M\,81 can be discussed in terms of a large number of
other nearby galaxies taken by the same facility.}

In this paper,
we aim {to point out the importance}
of FIR mapping data by deriving some fundamental
quantities about dust (dust temperature, dust optical
depth, etc.). At the same time, we try to clarify the
relation between global (integrated) quantities
and spatially resolved quantities. Some important
features in the FIR colour--colour diagram of a
spatially resolved
image is also focused on. The method used in this
paper is general and can be applied to future data
taken by \textit{Herschel}
or the Atacama Large Millimetre/submillimetre Array
(ALMA).\footnote{http://www.almaobservatory.org/}

This paper is organized as follows. We explain the data
analysis in Section \ref{sec:data}, and describe some
basic results related to global properties such
as morphology and radial dependence in
Section \ref{sec:result}. We investigate the
distribution of dust temperatures and the
colour--colour diagram
in Section \ref{sec:temperature}. We discuss the
results obtained for M\,81 in general contexts in
Section \ref{sec:discussion}.
Finally, Section \ref{sec:conclusion} gives the
conclusion.
The distance to M\,81 is assumed to be 3.63\,Mpc
\citep{freedman94} throughout
this paper. At this distance, 1\,arcmin corresponds
to 1.06\,kpc.

\section{Data analysis}\label{sec:data}

M\,81 was observed by \textit{AKARI}/FIS as a pointed
observation (PI: FIS Team). The data are taken from
Data Archives and Transmission
System.\footnote{http://darts.isas.jaxa.jp/}
Photometric observations were performed with four bands:
\textit{N}60 (central wavelength: 65\,$\micron$),
\textit{WIDE-S} (90\,$\micron$), \textit{WIDE-L}
(140\,$\micron$), and \textit{N}160 (160\,$\micron$)
with an observational mode of FIS01
(photometry/mapping mode), a reset interval of 0.5\,s and
a scan speed of 15 arcsec s$^{-1}$. We utilize the
images of the 65 $\micron$, 90 $\micron$, and
140 $\micron$ bands. The quality of the 160 $\micron$
band image is not good enough
for our purpose.
The FIS observation of M\,81 is composed of
two pointings: one covers almost all the M\,81 area
except for the south-east edge, which is observed by
the second observation. We combine those two images
to obtain the entire M\,81 image after the
background subtraction (Section \ref{subsec:bak}),
and the analysis procedures before combining the images
are identical for the two images.
The measured FWHMs of point spread
function are $37''$, $39''$, and $58''$ \citep{kawada07},
which correspond to 650 pc, 690 pc, and
1000 pc, respectively, at M\,81. Thus, internal structures
such as
spiral arms can be clearly identified on the images
(Section \ref{subsec:morphology}).


\subsection{FIS Slow-Scan Tool}

The raw data were reduced by using the FIS Slow-Scan
Tool (version 20070914;
\citealt*{verdugo07}).\footnote{http://www.ir.isas.ac.jp/ASTRO-F/Observation/}
The process includes flagging of bad data, measurement
of sky signal, dark and response correction,
flat-fielding, and construction of co-added
images. We used the local-flat for the flat-fielding.
The output image grid size is chosen to be 30'', which is 
about half of the beam size of 140 \,$\micron$. 
We confirmed that the following 
results are not sensitive to the selection of the grid
size.
{The final image is smoothed with a length of
$\sim 40''$ for the SW (65 $\micron$ and 90 $\micron$)
images and $\sim 60''$ for the LW (140 $\micron$) image to
avoid the effect of small scale detector noise.}

It is known that the \textit{AKARI}/FIS detectors
underestimate the total flux probably because of the
slow response \citep{shirahata08}. Thus, we multiply
the correction factors for the intensity, 1.7 for
65\,$\micron$ and 90\,$\micron$, and 1.9 for
140\,$\micron$, are multiplied to each band.
Because the factors are similar to all the bands, the colours
(flux ratios) and dust temperatures are not significantly
affected by this correction.

\subsection{Colour correction}\label{subsec:colorcorr}

Since the intensity at the central wavelength is
derived by assuming a flat spectrum
($\nu I_\nu =\mbox{constant}$) in the FIS Slow-Scan
Tool, {a} colour correction should be applied. We
applied correction to the \textit{WIDE-S}
(90 $\micron$) and
\textit{WIDE-L} (140 $\micron$) fluxes, assuming a
spectral shape of
$I_\nu\propto \nu^\beta B_\nu (T_\mathrm{d})$, where
$T_\mathrm{d}$ is the dust temperature, $\beta$ is the
emissivity index, and $B_\nu (T)$ is the Planck
function. The emissivity index $\beta$ is chosen to be 2
in this paper unless otherwise stated.
{The value of $\beta$ is dependent on the
grain composition, and $\beta =2$ is appropriate for
astronomical silicate
and graphite in \citet{draine84}.}
We apply
$T_\mathrm{d}=27$ K and 19 K for \textit{WIDE-S} and
\textit{WIDE-L}, respectively;
the former (latter) temperature
is derived from the total fluxes of M\,81 at \textit{N}60
and \textit{WIDE-S}
(\textit{WIDE-S} and \textit{WIDE-L})
(see Section \ref{subsec:global}).
Consequently,
the correction factor is 0.92 and 0.94 for the
\textit{WIDE-S} and \textit{WIDE-L} fluxes, respectively,
and the flux obtained by the
Slow Scan Tool is divided by these factors. The
uncertainty caused by the
colour correction (5 per cent for \textit{WIDE-S} and 1 per cent
for \textit{WIDE-L}
for the temperature range observed in the M\,81 disc;
Section \ref{subsec:Tmap}) is smaller than
the errors caused by the
background fluctuation.
For the narrow band \textit{N}60 (65 $\micron$) we do not apply
the correction,
since the colour correction only changes the flux by less
than 3 per cent.

\subsection{Background subtraction}
\label{subsec:bak}

Bright stripes along the scan direction caused by glitches
or non-uniform detector sensitivity are
observed in all the three bands.
In order
to eliminate these structures, the background levels are estimated
for each line along the scan direction
by averaging the intensities in the two sections 
before and after the scan of the M\,81 main body
with a separation of
about $18'$ and a total length of about $20'$, and then subtracted.
The RMSs of the background are 
1.3, 0.7, 2.9 MJy sr$^{-1}$ for the 65, 90, and
140 $\micron$ bands, respectively.
These values are estimated before eliminating the
stripes to show the original 
uncertainty including the non-uniform sensitivity 
and the time variability of the detector response.

\subsection{Matching the positions}

We matched the positions of the images taken by two
detectors, SW (\textit{N}60 and \textit{WIDE-S}) and LW
(\textit{WIDE-L}) {according to the position of 
central peak of the galaxy image.} The uncertainty in the 
relative position between
the detectors is well below the grid size.

\subsection{The physical quantities derived from the data}
\label{subsec:sedfit}

The intensity (surface brightness) at a wavelength
$\lambda$ (frequency $\nu\equiv c/\lambda$, where $c$
is the light speed) in each grid is denoted as
$I_\nu (\lambda )$. The intensity
ratio, $I_\nu (\lambda_1)/I_\nu (\lambda_2)$, is
called $\lambda_1-\lambda_2$ colour in this paper.
We derive the dust temperature $T_\mathrm{d}$ in
each grid as follows.
{Since the dust emission is
optically thin in the FIR (Section \ref{subsec:Tmap}),
$I_\nu (\lambda)=\tau (\lambda )B_\nu (T_\mathrm{d})$,
where $\tau (\lambda )$ is the optical depth at
wavelength $\lambda$. The functional form of the
optical depth is assumed to be
$\tau (\lambda )=A\nu^2$ ($\nu =c/\lambda$)
with an unknown constant $A$,
which is independent of the frequency.
This functional form for the optical depth
indicates that the frequency dependence
of the dust emission coefficient follows
$\propto \nu^\beta$ with
$\beta =2$.\footnote{Although
we consider a variation of $\beta$ as a function
of wavelength in
Section \ref{subsec:clr} for a detailed theoretical
model, the simple and common assumption of $\beta =2$ is
useful to compare our observational results with
the quantities obtained in the literature.}
Then, with two wavelengths
($\lambda_1$ and $\lambda_2$) selected, a set of
equations,
$I_{\nu}(\lambda_1)=\tau (\lambda_1 )
B_\nu (T_\mathrm{d})|_{\lambda_1}$
and $I_\nu(\lambda_2)=\tau (\lambda_2)
B_\nu (T_\mathrm{d})|_{\lambda_2}$,
is solved to obtain $T_\mathrm{d}$ and $A$. The dust
temperature $T_\mathrm{d}$ determined from
wavelengths of 90\,$\micron$ and 140\,$\micron$
represents the temperature of large grains in radiative
equilibrium with the ambient radiation field
\citep[e.g.][]{li01},
and it is denoted as $T_\mathrm{LG}$.}

It is convenient to convert $\tau (\lambda)=A\nu^2$ to a
commonly used indicator of dust optical depth. We choose
$A_V$ (the extinction in $V$ band in units of magnitude)
for such an indicator. We adopt the Galactic extinction
curve for the conversion from $\tau (\lambda )$ to
$A_V$ as derived by \citet{weingartner01} for $R_V=3.1$:
$A_V=\mathcal{C}(\lambda )\tau (\lambda )$,
where the factor $\mathcal{C}(\lambda )$ is $8.3\times 10^2$
for 90\,$\micron$ and $1.9\times 10^3$ for 140\,$\micron$.
Although we derive $A_V$ by using the 90\,$\micron$ values,
we obtain the same value for $A_V$ also from the 140\,$\micron$
values because
$\mathcal{C}(\lambda )\propto\lambda^{2}$ holds within
10 per cent between 90 and 140\,$\micron$.

\section{Global Properties}\label{sec:result}

\begin{figure}
\includegraphics[width=0.45\textwidth]{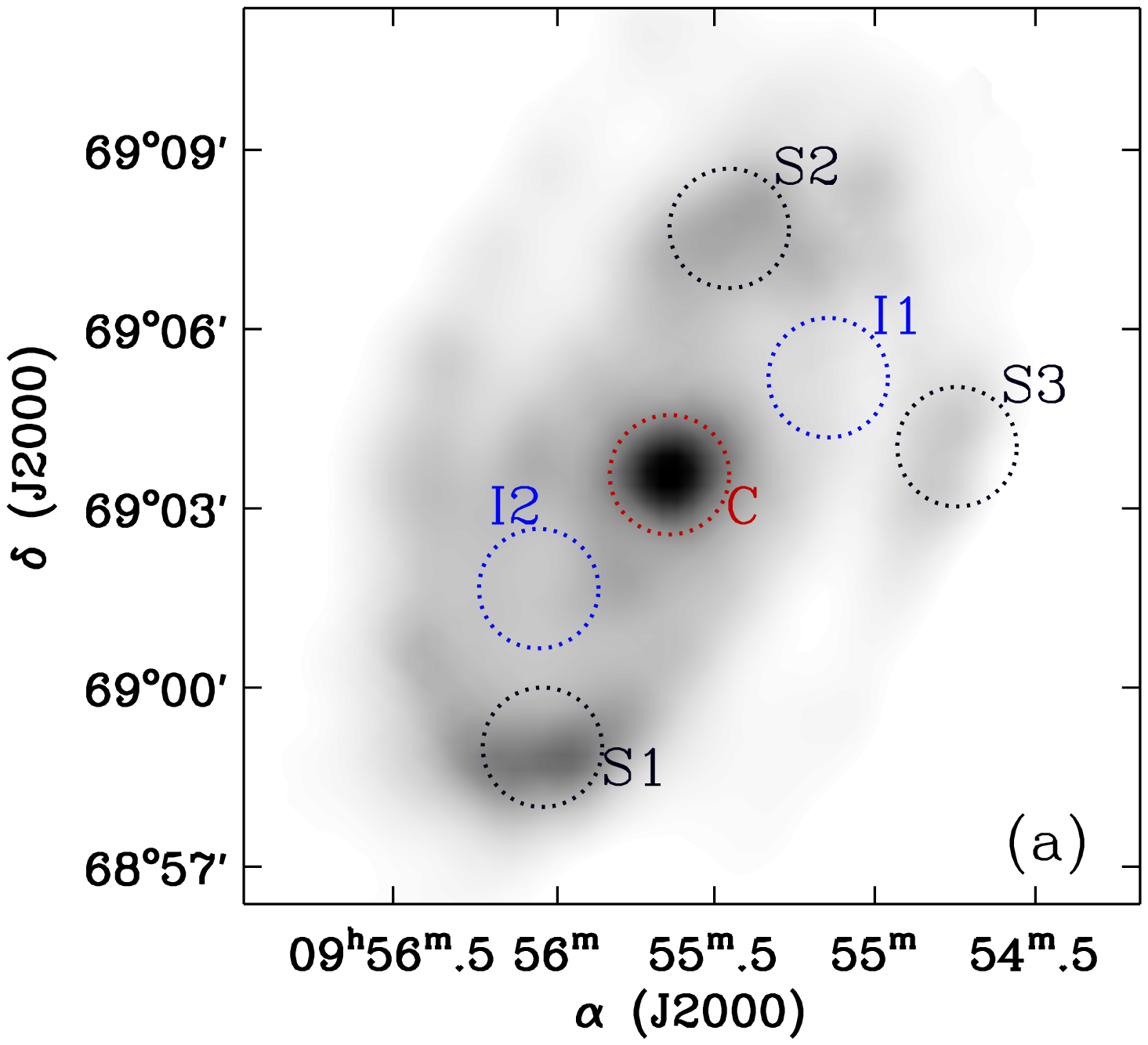}
\includegraphics[width=0.45\textwidth]{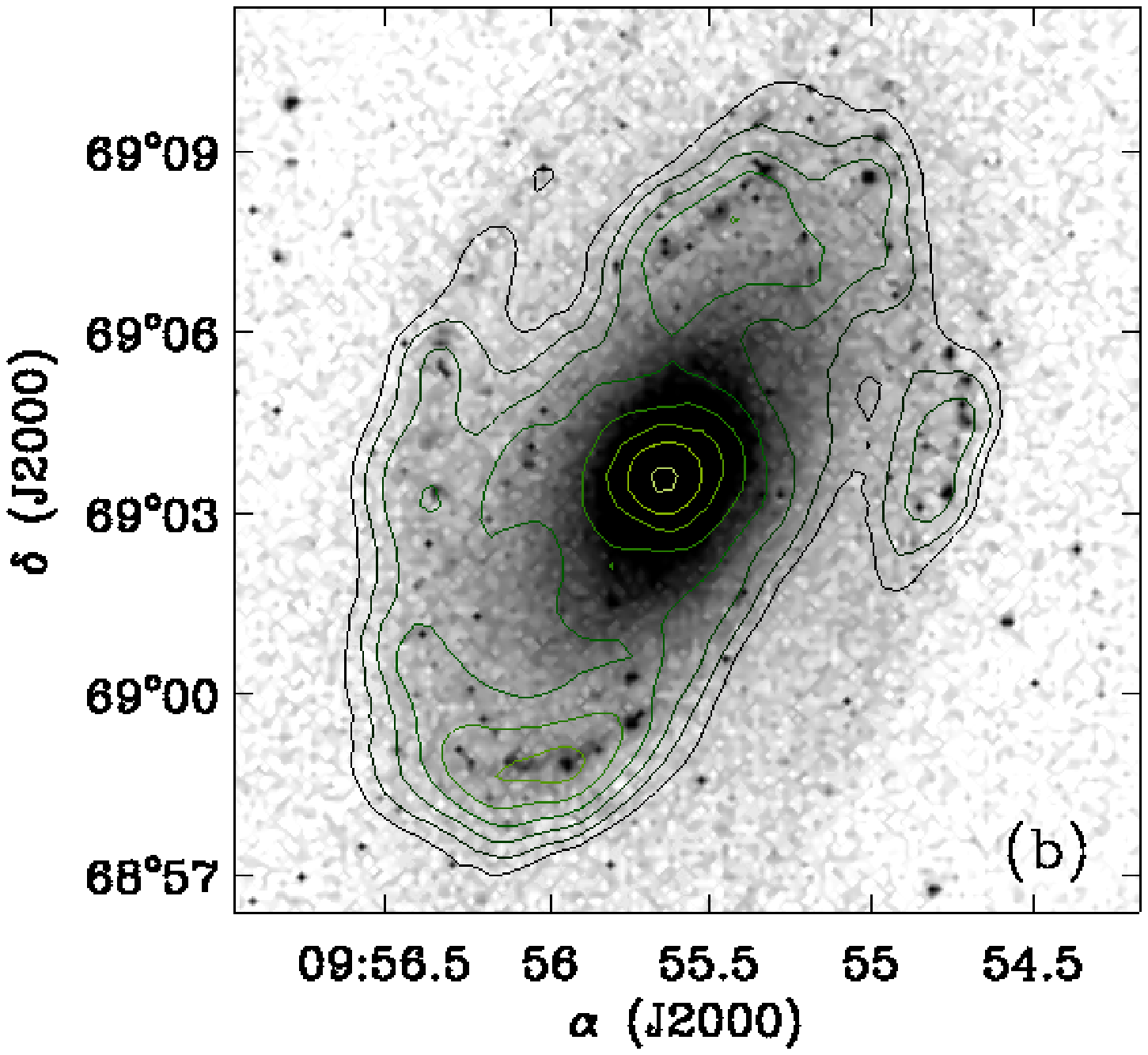}
 \caption{Images of M\,81. (a) The 90 $\micron$ image.
 The circles define various regions used for the
analysis of the colour--colour relation in
Section \ref{subsec:clr}: the central region (C), the spiral
arm regions (S1, S2, and S3), and the interarm regions (I1,
and I2). (b) The H$\alpha$ image taken by \citet{cheng97}
(grey scale) overlaid with the 90 $\micron$ brightness
contours. The highest
contour shows 0.71 times the peak intensity and the
intensity ratio of two adjacent contours is set to be
0.71.}
 \label{fig:image}
\end{figure}

\subsection{Morphological features}\label{subsec:morphology}

In Fig.\ \ref{fig:image}, we show the 90 $\micron$ image.
The central bright peak, the spiral structures, and the
spots along the spiral arms are clear. In order to examine
the correspondence between the FIR brightness and the star
formation activity in M\,81, we compare our FIR image with
the H$\alpha$ image taken from the NASA/IPAC Extragalactic
Database (NED)\footnote{http://nedwww.ipac.caltech.edu}
(originally from \citealt{cheng97}). The FIR emission really
traces the H$\alpha$ central peak, the H$\alpha$ spiral
arms, the H\,\textsc{ii} regions along the spiral arms.
This confirms that the FIR emission is a good tracer
of the star formation activities in galaxies.

\subsection{Spectral Energy Distribution}
\label{subsec:global}

\begin{figure}
\includegraphics[width=0.45\textwidth]{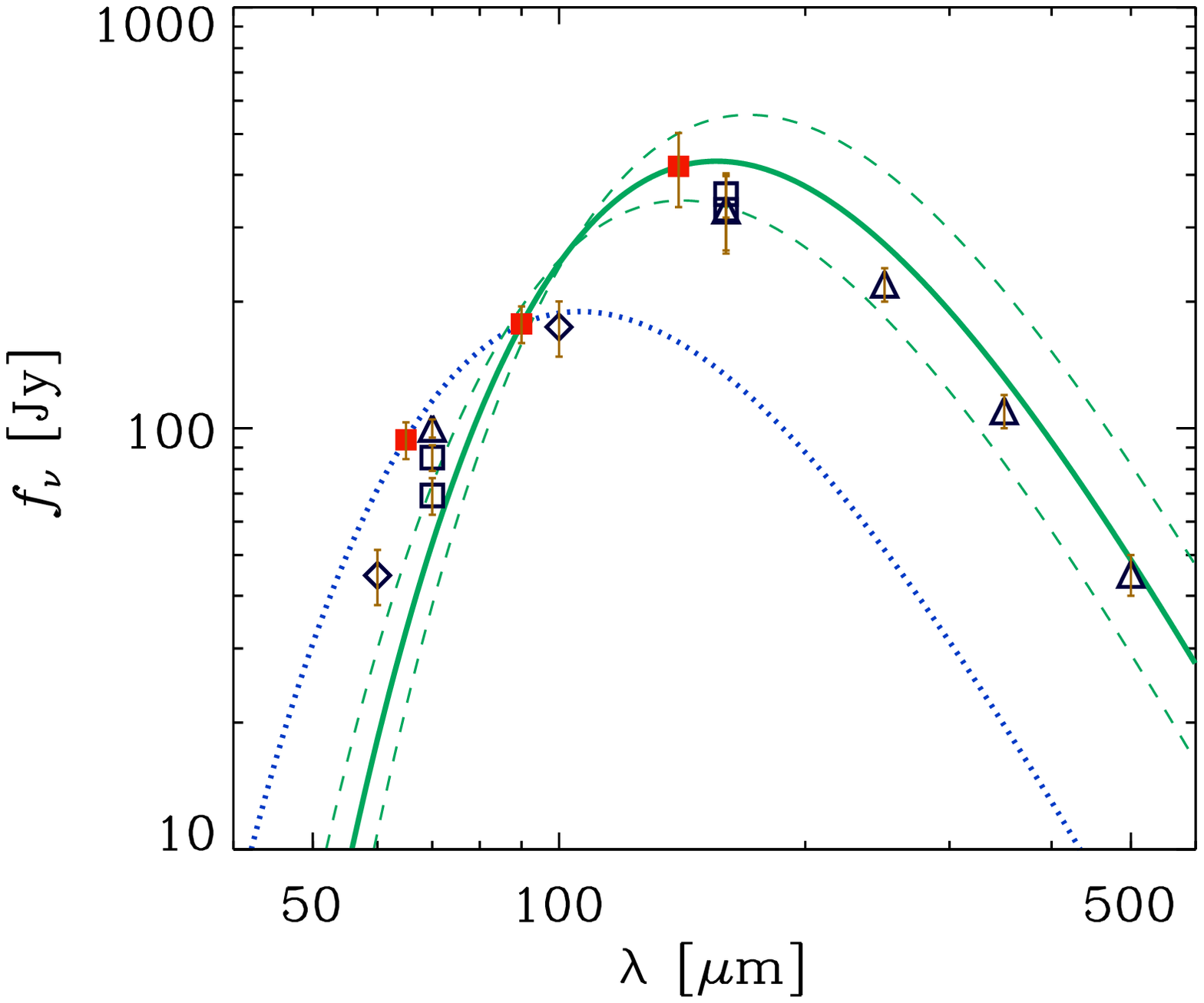}
\caption{FIR SED of M\,81. The filled squares show our
measurements, while the others are taken from the
literature: The open diamonds at 60\,$\micron$ and
100\,$\micron$ are taken from the \textit{IRAS}
data \citep{rice88}, the open squares at 70\,$\micron$
and 160\,$\micron$ from the \textit{Spitzer} data
\citep{perez06,dale07}, {and the open triangles from the 
\textit{Herschel} data \citep{bendo10}}. The solid (dotted) line shows
the single-temperature SED which explains our
measurements at 90\,$\micron$ and 140\,$\micron$
(65\,$\micron$ and 90\,$\micron$).
The dashed lines show the SEDs at the 
temperature of upper and lower bound of $T_\mathrm{LG}$ estimated by
using the upper and lower fluxes at 90\,$\micron$ and 140\,$\micron$.
}
 \label{fig:sed}
\end{figure}

The measured total fluxes, $f_\nu$, at 65, 90, and
140\,$\micron$
are shown in Fig.~\ref{fig:sed} and Table~\ref{tab:flux}
together with the data taken from the literature.
{The fluxes are estimated by integrating intensities 
over a rectangular region of width $15'$ and length $18'$ 
covering the galaxy main body.}
We conservatively adopt flux errors
of 10 per cent for the 65\,$\micron$ band
and the 90\,$\micron$ band and 20 per cent for the
140\,$\micron$ band \citep{hirashita08}.
{
As shown in Table \ref{tab:flux}, we obtain the
fluxes consistent with \citet{suzuki10} within the errors.
}

\begin{table}
\centering
\begin{minipage}{80mm}
\caption{Observational fluxes in units of Jy for the entire
M\,81 (global) and
the individual regions shown in Fig.\ \ref{fig:image}.}
\label{tab:flux}
    \begin{tabular}{lccc}
     \hline
     Wavelength & 65 $\micron$ & 90 $\micron$ & 140 $\micron$\\ 
     \hline 
     Global & 94 $\pm$ 9 & 177 $\pm$ 18 & 419 $\pm$ 84 \\
     Global (S10)\,$^\mathrm{a}$ & $70\pm 20$ & $189\pm 40$
       & $356\pm 140$ \\
     C  & 9.21 & 12.8 & 16.3 \\
     S1 & 3.61 & 6.05 & 10.7 \\
     S2 & 3.42 & 6.52 & 13.0 \\
     S3 & 1.65 & 2.71 & 5.32 \\
     I1 & 2.06 & 4.80 & 10.8 \\
     I2 & 2.12 & 5.07 & 11.7 \\
     \hline
    \end{tabular}

\medskip

$^\mathrm{a}$ Measurements by \citet{suzuki10}.\\
\textit{Note.} The conservative errors are 10 per cent for
65 and 90 $\micron$, and 20 per cent for 140 $\micron$.
\end{minipage}
\end{table}

We also show two spectral energy distributions (SEDs)
which fit the data at 90 and 140\,$\micron$ and
those at 65 and 90\,$\micron$ with the functional form
given in Section \ref{subsec:sedfit} (i.e.\
$f_\nu\propto\nu^2B_\nu (T_\mathrm{d})$). We observe that
the SED which fits the 90\,$\micron$ and 140\,$\micron$
fluxes underproduces the 65\,$\micron$
flux. This indicates the contribution of very small
grains (VSGs) to the emission at $\la 65\,\micron$
\citep{aannestad79,draine85}: Small grains with radii
$\la 10$\,nm
are stochastically (i.e.\ not in radiative
equilibrium) heated to a high temperature,
contributing to the intensity at the short
($\la 65\,\micron$) wavelengths.
The dust temperature $T_\mathrm{d}$ derived from the
90\,$\micron$ and 140\,$\micron$ fluxes is 18.6\,K,
while that derived from the 65\,$\micron$ and 90\,$\micron$ is
27.3\,K. 
The former temperature is interpreted to be the one
of large grains which are in radiative equilibrium with
the ambient ISRF ($T_\mathrm{LG}$), while the latter does
not present any physical dust temperature because of the
contamination of stochastically heated very small grains
to the 65 $\micron$ flux. To estimate the uncertainty
in $T_\mathrm{LG}$, we use the upper (lower) and lower
(upper) values for the 90 and 140 $\micron$ fluxes
and obtain the upper (lower) temperature as shown in
Fig.\ \ref{fig:sed}: 20.5 and 17.0 K
for the upper and lower temperatures, respectively.
Thus, we finally
obtain an estimate for the large grain temperature
as $T_\mathrm{LG}=18.6^{+1.9}_{-1.6}$ K from
the global SED. We observe that the \textit{Spitzer} and
\textit{Herschel} data at $\lambda\geq 160~\micron$ are
broadly consistent with our
SED fit to the
90 $\micron$ and 140 $\micron$ data within the errors.

\citet{perez06} have shown that the \textit{Spitzer} data
can be fitted with two dust temperatures: their colder
component with a temperature of $18\pm 2$ K has a
consistent temperature with our $T_\mathrm{LG}$, while
their warmer component with $53\pm 7$ K shows a
much higher temperature because they also used the
24 $\micron$ data.
The temperature estimated by the \textit{Herschel} data is
$17.8\pm 0.6$ K \citep{bendo10}.
Our estimate of the large grain temperature is consistent
with the \textit{Spitzer} and \textit{Herschel} results
within the errors.

As a global quantity, we derive the total infrared (TIR)
luminosity (the integrated luminosity
between 8 and 1000 $\micron$), which is used later in
Section \ref{subsec:Tmap}.
For this aim, we first estimate
the total flux traced by \textit{AKARI} FIS.
The \textit{AKARI} FIR flux, $f_\mathit{AKARI}$ is defined by
\citet{hirashita08} to estimate the total flux in the
wavelength range of 50--170 $\micron$:
\begin{eqnarray}
f_\mathit{AKARI} & \equiv & f_\nu (65~\micron )\Delta\nu
(\mbox{\textit{N}60})+f_\nu (90~\micron )\Delta\nu
(\mbox{\textit{WIDE-S}})\nonumber\\
& &
+f_\nu (140~\micron )\Delta\nu(\mbox{\textit{WIDE-L}}),
\end{eqnarray}
where $f_\nu (\lambda )$ is the flux per unit frequency at
wavelength $\lambda$, and $\Delta\nu (\mbox{band})$ denotes
the frequency width covered by the band. According to
\citet{kawada07}, we adopt $\Delta\nu (\mbox{N60})=1.58$ THz,
$\Delta\nu (\mbox{\textit{WIDE-S}})=1.47$ THz, and
$\Delta\nu (\mbox{\textit{WIDE-L}})=0.831$ THz.
With $f_\mathit{AKARI}$, the \textit{AKARI} FIR luminosity,
$L_\mathit{AKARI}$, is estimated as
\begin{eqnarray}
L_\mathit{AKARI}=4\pi D^2f_\mathit{AKARI},
\end{eqnarray}
where $D=3.63$ Mpc is the distance to M\,81.
We obtain
$L_\mathit{AKARI}=(3.1\pm 0.6)\times 10^9~\mathrm{L}_{\sun}$.
Considering that the 10--20 per cent errors in the
flux, we put an error of 20 per cent also to the
luminosity.

Next, we relate the \textit{AKARI} luminosity and
the TIR luminosity. There is a good correlation
between $L_\mathit{AKARI}$ and $L_\mathrm{TIR}$ (total infrared
luminosity emitted by dust) in the \textit{AKARI} FIS
Bright Source Catalogue, the first
primary catalogue
from the \textit{AKARI} All-Sky Survey \citep{yamamura10}, as shown
by \citet{takeuchi10}:
\begin{eqnarray}
\log L_\mathrm{TIR}=0.940\log L_\mathit{AKARI}+0.914.
\end{eqnarray}
With this empirical formula,
$L_\mathrm{TIR}=(6.9\pm 1.4)\times 10^{9}~\mathrm{L}_{\sun}$
for
M\,81. This luminosity is divided by $4\pi D^2$ to obtain
the total infrared flux,
$F_\mathrm{TIR}=1.7\times 10^{-8}$ erg cm$^{-2}$ s$^{-1}$.
Considering that the flux may be uncertain by 20 per cent,
it would be appropriate to put a 20 per cent error for
each flux or luminosity. \citet{suzuki10} obtained
smaller luminosity
$(4.0\pm 0.2)\times 10^9$ L$_{\sun}$ for the total
FIR luminosity.
They consider contributions from two components:
a cold component with a temperature of 22 K and a warm
component with a temperature of 64 K, but these
two components do not include the
contribution from the emission at $<20~\micron$.

We can also check the consistency of the dust mass
with the \textit{Herschel} data.
The total dust mass, $M_\mathrm{d}$, is estimated as
\begin{eqnarray}
M_\mathrm{d}=
\frac{f_\nu (\lambda)D^2}{\kappa_\nu B_\nu (T_\mathrm{d})},
\end{eqnarray}
where $\kappa_\nu$ is the mass absorption coefficient at
a frequency $\nu$.
{Using the measured flux at 90 $\micron$,
the dust temperature derived from the 90 and 140 $\micron$
data, and
$\kappa_\nu =34$ cm$^2$ g$^{-1}$ at 90 $\micron$
\citep{weingartner01},
we obtain
$M_\mathrm{d}=(3.2\pm 0.6)\times 10^7~\mathrm{M}_{\sun}$,
where an error of 20 per cent is applied because of the
flux uncertainty.
This value is consistent with the dust mass
($3.4\times 10^7~\mathrm{M}_{\sun}$) obtained
by the \textit{Herschel} observation \citep{bendo10}.
}

\subsection{Radial dependence}\label{subsec:radial}

\begin{figure}
\includegraphics[width=0.45\textwidth]{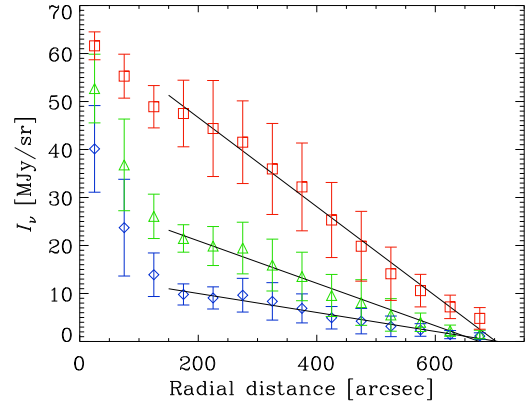}
 \caption{
Intensity profiles in various wavelengths. The diamonds,
triangles, and squares represent the averaged
intensities as a function of deprojected radial distance
in the 65, 90, and 140 $\micron$ bands, respectively. The
bars on the individual data points represent the standard
deviation, not the errors. The solid lines are the fitting
results with linear slopes for $r>150$ arcsec.
} \label{fig:radial}
\end{figure}

In Fig.\ \ref{fig:radial}, we show the radial profiles in
the three bands. The radial distances ($r$) are
deprojected by considering the position angle (157 degree)
and the inclination angle (57 degree) taken from NED.
The inclination
angles given in NED range from 55 to 63 degree.
Therefore the deprojected radius
is uncertain by 10 per cent.
The radius from the centre
is divided into bins with
a width of 50 arcsec, and the
intensities of the pixels contained in each radius bin is
averaged to obtain the intensity as a function of radius.
The standard deviation for each radius bin is also shown
{with the bar}.
The flux errors are much smaller than the standard deviations.

The radial profile can be divided into two components:
bulge ($r<150$ arcsec) and disc ($r\geq 150$ arcsec).
The 65 and 90\,$\micron$ band images show a bright
bulge where the intensity
increases rapidly as the radius decreases, while the
140 $\micron$ image does not have such a sharp rise.
The difference in the `sharpness' of the bulge component
among different bands is due to the high dust temperatures,
which reflect an intense radiation from the active
galactic nucleus (AGN) or from the stars with high
surface density. The bulge component is not prominent
also in the long-wavelength \textit{Herschel} SPIRE
data \citep{sauvage10}.

For the disk part, all the intensities in the three bands
show a linear slope along the radius. The linear slope
means that the dust
emission is radially more extended than the stellar
emission, whose radial profile drops exponentially
\citep[e.g.][]{baggett98}.
Although \citet{sauvage10}
apply a fitting with exponential or Gaussian function
to the FIR radial profile of the \textit{Herschel} data,
the fit is not necessarily good.
The dispersion shown by the bars is relatively large
around $r\sim 400$ arcsec. This radius corresponds to
the radial extent of the spiral arms.

\begin{figure}
\includegraphics[width=0.45\textwidth]{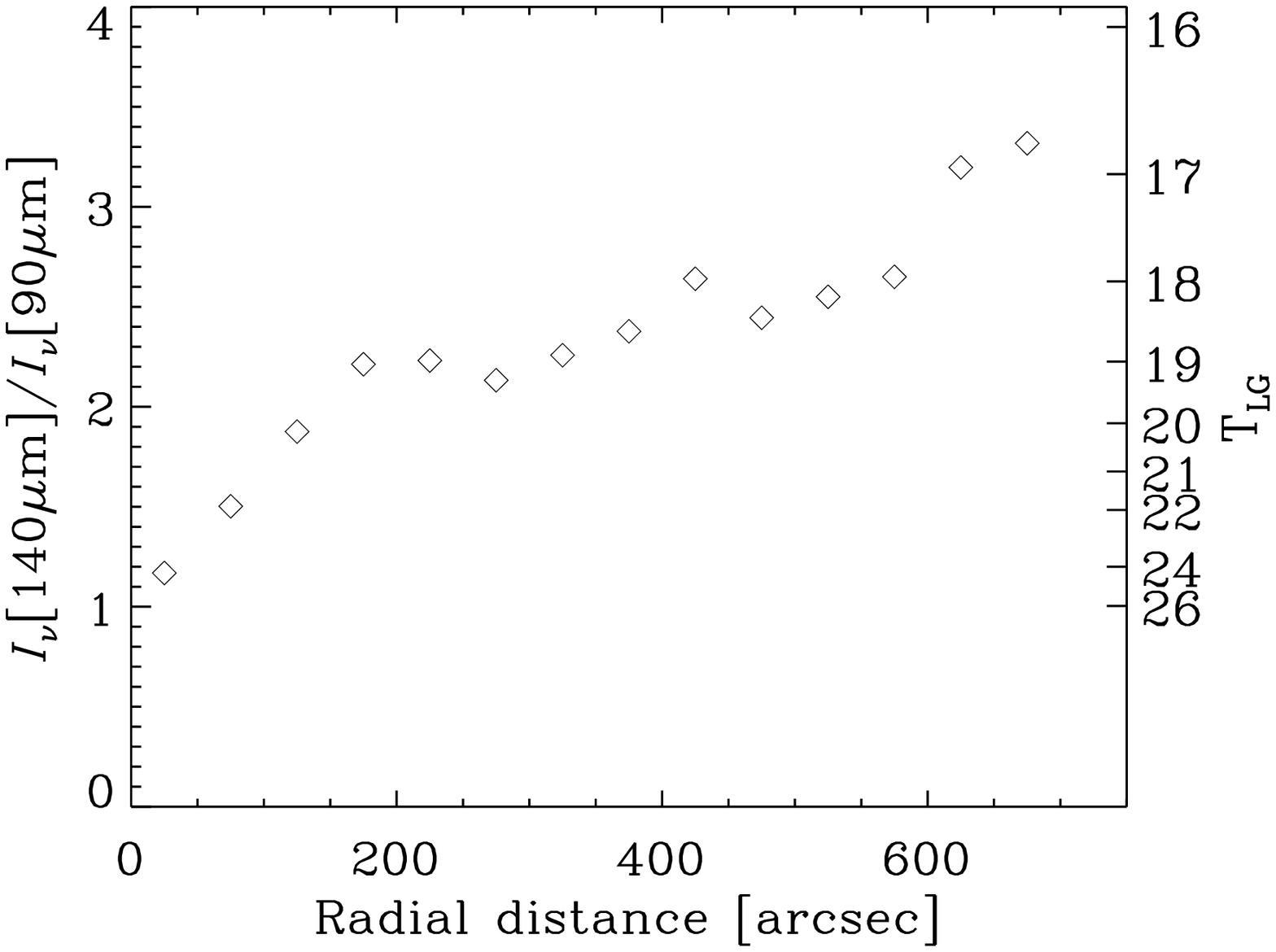}
\includegraphics[width=0.45\textwidth]{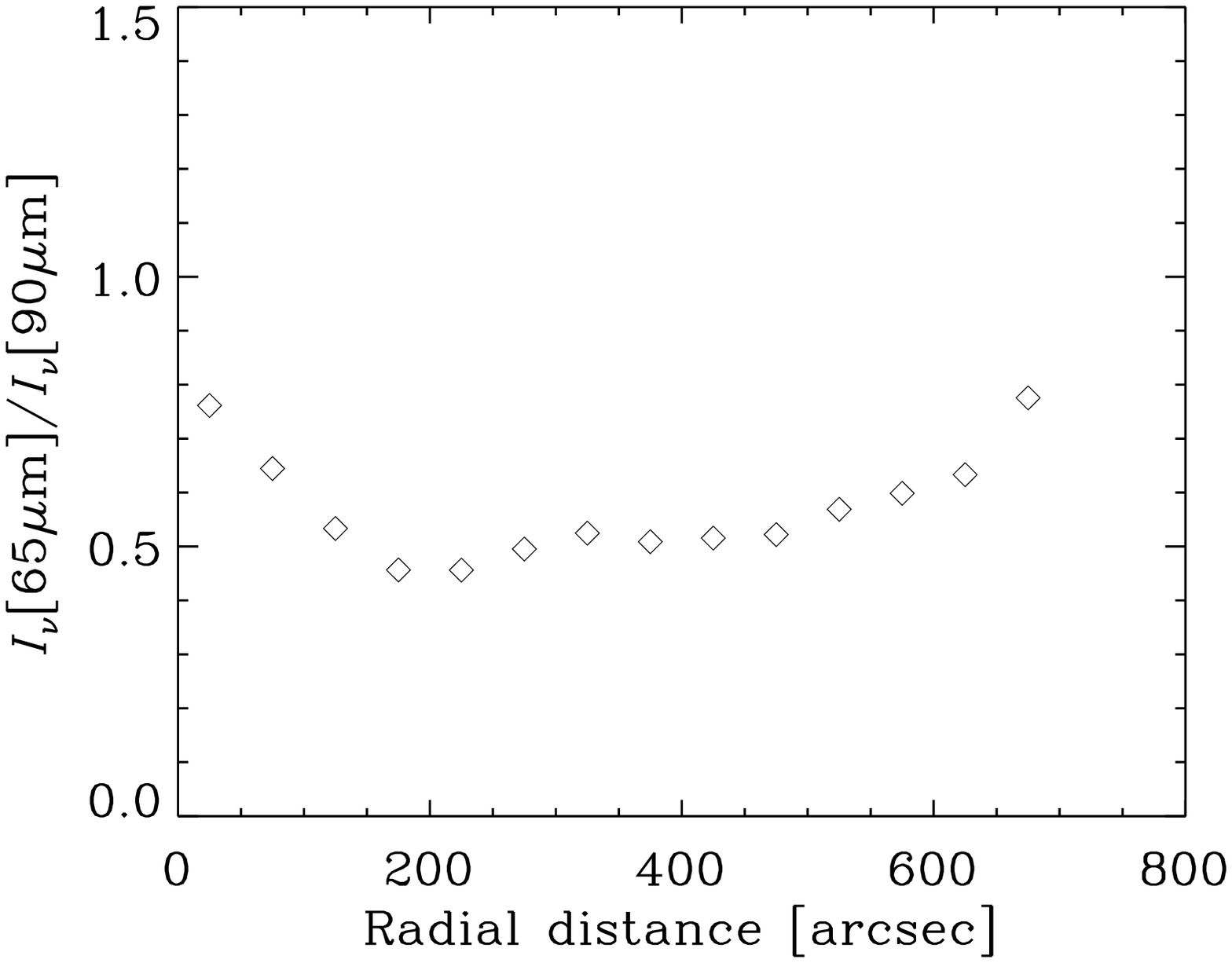}
 \caption{
Radial profiles
of the $140~\micron -90~\micron$ colour (upper)
and the $65~\micron -90~\micron$ colour (lower).
In the upper figure, the colour is converted into
the equilibrium dust temperature $T_\mathrm{LG}$
by assuming $\propto\nu^2B_\nu (T_\mathrm{LG})$
for the SED shape.
} \label{fig:radial_clr}
\end{figure}

To compare the radial extents in the three bands, we
show the radial profiles of the colours,
$I_\nu (140~\micron )/I_\nu (90~\micron)$ and
$I_\nu (65~\micron )/I_\nu (90~\micron)$ in
Fig.\ \ref{fig:radial_clr}. The former rises as the radius
increases, {showing that the 140 \,$\micron$ image 
is more extended than 90 \,$\micron$}, while the latter
dose not show such a
clear trend. The slight enhancement of the
$65\,\micron -140\,\micron$ colour and the plateau of
the $140\,\micron -90\,\micron$ colour
around $r=6$ arcsec are caused by the relatively
high temperature in the spiral arms.

Because of the linear-slope behaviour of the radial
profile, the radial extent is well defined by the
intercepts on the $x$ axis in Fig.~\ref{fig:radial}.
The radial extents defined in this way are
706, 674, and 702 arcsec for the 65, 90, and 140
$\micron$ images, respectively. The reason why the emission
is more extended in
140\,$\micron$ than in 90\,$\micron$ can be the radial gradient
of $T_\mathrm{LG}$. The $140\,\micron -90\,\micron$ colour
is directly converted into $T_\mathrm{LG}$ by assuming
a functional form of $\nu^2B_\nu (T_\mathrm{LG})$ for the
SED.
While 140\,$\micron$ is
closer to the intensity maximum of the SED,
90\,$\micron$ at the Wien side is more sensitive to the
temperature change, and decreases more rapidly outwards.
A positive $160\,\micron -70\,\micron$
colour gradient
and a negative temperature gradient of cold dust component 
are also shown with the \textit{Spitzer} data by \citet{perez06}.

{
\citet{bendo10} have also shown the radial profile of
the $70~\micron -160~\micron$ colour by using the
\textit{Herschel} data. Their results do not
show a monotonic decrease of the the
$70~\micron -160~\micron$ colour as a function of radius.
The non-monotonic behaviour is similar to our \textit{AKARI}
$65~\micron -90~\micron$ colour. On the other hand,
the \textit{Herschel} colours composed of the wavelengths
longer than $160~\micron$ show a monotonic trend, which
can be interpreted as a negative temperature gradient.
This is consistent with our $140~\micron -90~\micron$
colour. Therefore, if a wavelength shorter than
$\sim 70~\micron$ is included, the colour behaves in a more
complicated way possibly due to a relatively strong
contribution from the local heating source to the short
wavelength range.
}



\section{Detailed Features}\label{sec:temperature}

\subsection{Distribution of dust temperature and
optical depth}\label{subsec:Tmap}

\begin{figure}
\includegraphics[width=0.45\textwidth]{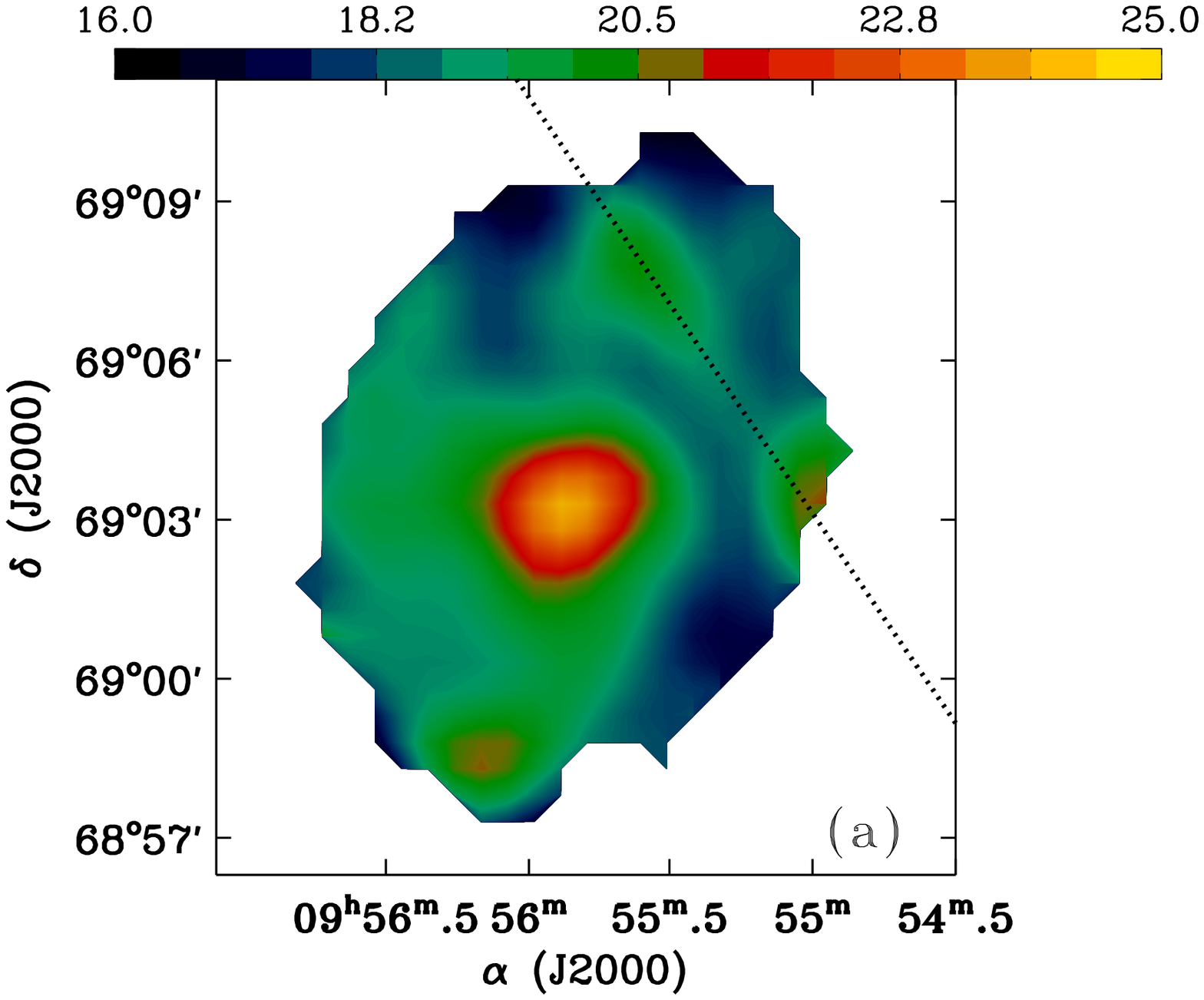}
\includegraphics[width=0.45\textwidth]{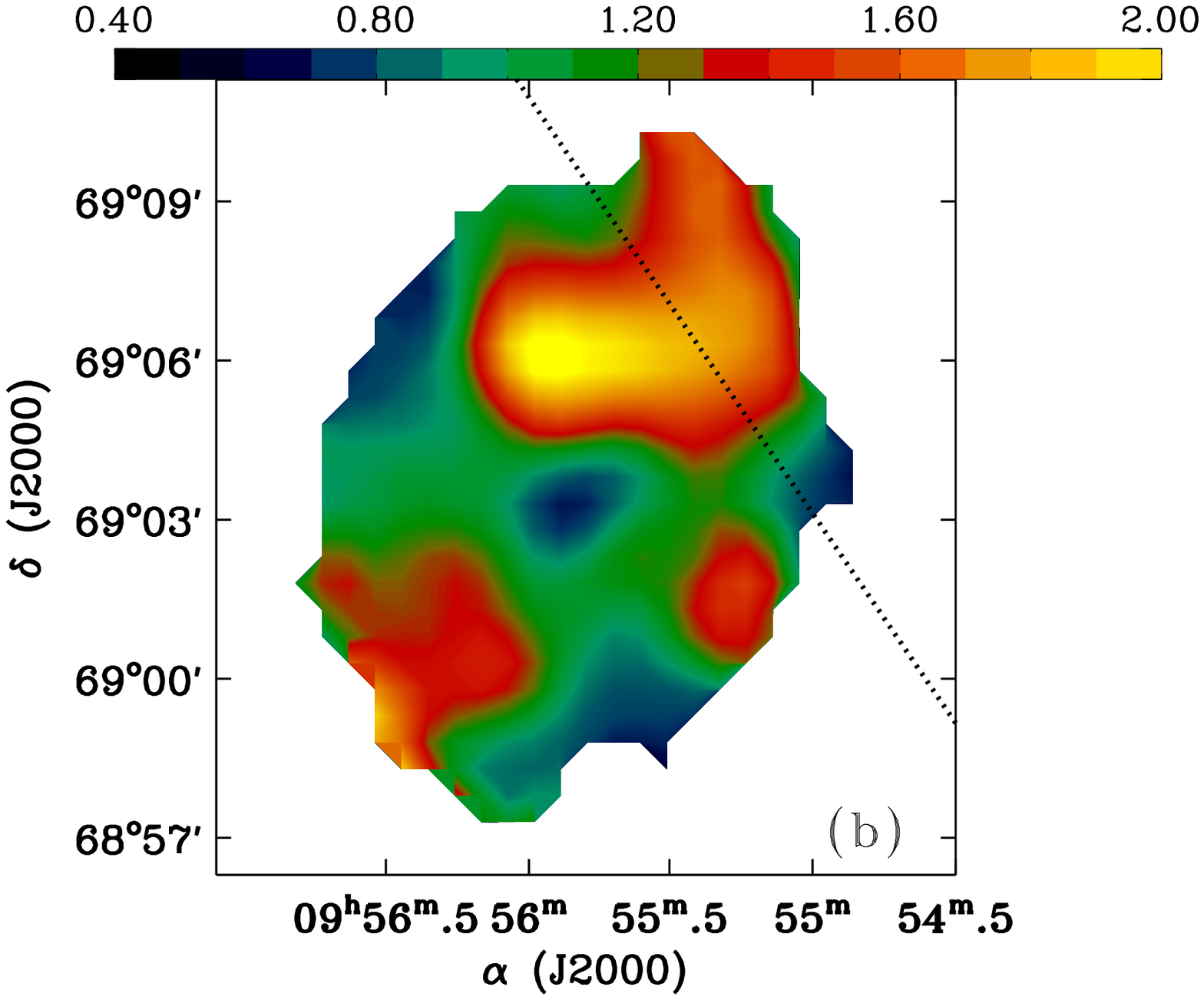}
 \caption{Spatial distribution of (a) large-grain
temperature ($T_\mathrm{LG}$) and (b) dust extinction in
the $V$ band ($A_V$). Both quantities are estimated from
the 90 $\micron$ and 140 $\micron$ intensities in the
individual grids. The correspondence between the
colours and the physical quantities is shown in the
colour bar in each panel.
The dotted lines indicate the most prominent stripe 
and the scan direction.}
 \label{fig:dist_T_tau}
\end{figure}

The large-grain temperature ($T_\mathrm{LG}$) is
determined at each grid by the method described in
Section \ref{subsec:sedfit}.
{We only adopt the 
pixels with brightnesses above 4 $\sigma$ of the background
fluctuation, that is 
$I_\nu (90~\micron )>2.8~\mathrm{MJy~sr}^{-1}$
and $I_\nu (140~\micron )>11.6~\mathrm{MJy~sr}^{-1}$ for
the temperature map to avoid artificial
features.}
The result is shown in Fig.\ \ref{fig:dist_T_tau}.
The spatial distribution of $T_\mathrm{LG}$ shows some
structures in M\,81, especially the central peak and
the warm spots along the spiral arms. {There is an
overall rough trend that the temperature decreases outwards,
which is consistent with the colour gradient shown in
Section \ref{subsec:radial}.}
The temperature is typically {25} K at the centre,
{20--21} K at
the warm spots along the arms, and {17--20} K in the other
diffuse regions, indicating more intense radiation field in
the centre and in the spots. Moreover, the dust temperature
derived from the total fluxes ({18.6} K;
Section \ref{subsec:global}) is consistent with
that in the diffuse regions, which supports the view
that diffuse (general) ISRF is responsible for the major part
of dust heating \citep{walterbos96}. 
{The diffuse nature of the cool
($T_\mathrm{LG}\simeq 18$ K) dust is consistent with the
Spitzer 160 $\micron$ image \citep{perez06}.}

We also present the spatial distribution of dust optical
depth in Fig.\ \ref{fig:dist_T_tau}. The optical depth
at 90 $\micron$ is converted to the $V$-band extinction
($A_V$) (Section \ref{subsec:sedfit}). There is a weak
anti-correlation between $A_V$ and $T_\mathrm{LG}$. 
The large excess of $A_V$ in the north-west and south-west
parts of the disc may be strongly affected by the contamination of
stripes caused by unstable detector response.
In Fig.\ \ref{fig:dist_T_tau}, the most prominent 
stripe is shown by the dotted lines
indicating the scan direction. In the temperature map, there are
two high-temperature islands connected with a bridge on this stripe.
The neighbouring region around this line with a width of 3 arcmin may
not be reliable because of some stripes.
Thus, future observations of the structures around the
stripes are required.

Our estimate of $A_V$ is reasonable if the dust in each pixel
has an uniform
temperature. However, in the spiral arms, as shown in the
H$\alpha$ image (Fig.\ \ref{fig:image}), star-forming
regions whose sizes are smaller than the grid size of the
FIR image are clearly seen. If the dust temperature is
biased to these regions, the dust optical depth can be
underestimated. In fact, as shown later in Section
\ref{subsec:clr}, the FIR colour--colour relation also
supports a significant contribution of warm star-forming
regions to the FIR emission in spiral arms.
The central part does not show a strong
indication of such inhomogeneous radiation field;
that is, the dust temperature in that
region might be well approximated by a single component.
Thus, there is still a possibility that small $A_V$ in the
central part is due to
a deficiency of dust. However, since $A_V$ is very sensitive to
the determination of the dust temperature and our results are 
affected by unstable detector response, all the
features in $A_V$ should be confirmed by future observations.

The $A_V$ value is roughly
{0.5--2} in the M\,81 disc.
\citet{buat05} show that FIR to ultraviolet (UV) luminosity
ratio of galaxies is related to the dust extinction because
the efficiency of UV reprocessing into FIR by dust is
related to the dust extinction. The UV flux of M\,81 at the
\textit{FUV} band (1528 \AA) of
the \textit{Galaxy Evolution Explorer} (\textit{GALEX}) is
$F_\nu =179$ mJy \citep{dale07}. Thus,
$F_\mathit{FUV}\equiv\nu F_\nu =3.51\times 10^{-9}$
erg cm$^{-2}$ s$^{-1}$. The extinction at the \textit{FUV}
band can be estimated by using the fitting formula
derived from the model calculations by \citet{buat05}:
\begin{eqnarray}
A_\mathit{FUV}=-0.0333y^3+0.3522y^2+1.1960y+0.4967,
\end{eqnarray}
where $y\equiv\log (F_\mathrm{TIR}/F_\mathit{FUV})$. Using
the value for $F_\mathrm{TIR}$ in
Section \ref{subsec:global},
we obtain $A_\mathit{FUV}=1.46$. This matches the mean
value for the UV-selected sample in \citet{buat05} in the
local Universe. Now using the Galactic
extinction curve given by \citet{weingartner01}
with $R_V=3.1$ ($A_V/A_\mathit{FUV}=0.389$),
we obtain $A_V=0.57$. 
The extinction shown in Fig.\ \ref{fig:dist_T_tau} is
the one over the entire column in the galactic
disc. If the stars are on average located in the
middle of the disc thickness, the stellar extinction
would be half of the values obtained in
Fig.\ \ref{fig:dist_T_tau}; that is, $A_V\simeq 0.25$--1.
$A_V=0.57$ is within this range.

\subsection{Colour--colour diagram}\label{subsec:clr}

\begin{figure}
\includegraphics[width=0.45\textwidth]{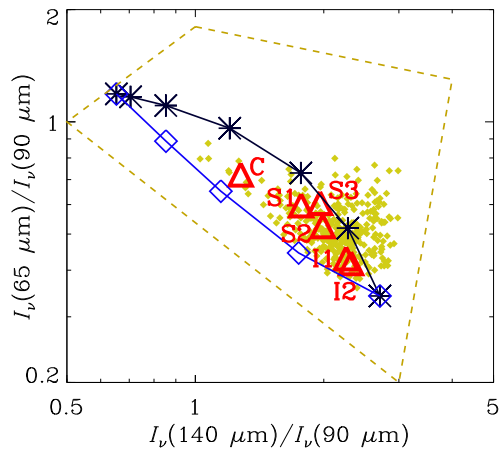}
 \caption{The relation between $65~\micron -90~\micron$
 colour and $140~\micron -90~\micron$ colour. Small filled
 diamonds show the colours in the individual grids in
 M\,81. Triangles represents the colours for the flux
 integrated in the regions depicted in
 Fig.\ \ref{fig:image}. Large
 open diamonds present the theoretical colour--colour
 relation with $\chi =1$, 3, 10, 30, and 100, and asterisks
 show the results with $f_\mathrm{h}=0$, $10^{-2.5}$,
 $10^{-2}$, $10^{-1.5}$,
 $10^{-1}$, $10^{-0.5}$, and 1 from
 large to small $I_\nu (140~\micron )/I_\nu (90~\micron )$.
 The area shown by the large square with
 dashed lines show the representative colour--colour
 relations of the galaxy sample in the \textit{AKARI} FIS
 Bright Source Catalogue
 \citep{pollo10}.
 }
 \label{fig:clr}
\end{figure}

The relation among the intensities at the three bands
is investigated here. Following \citet{hibi06}, we
investigate the FIR colour--colour relation.
Fig.\ \ref{fig:clr} shows the relation between the
$65~\micron -90~\micron$ colour
and the $140~\micron -90~\micron$ colour
{for the individual grids whose intensities are
above 4 $\sigma$ of the background fluctuation
for all the bands (i.e., the same criterion for
the temperature map in Section \ref{subsec:Tmap})}.
The overall
trend from the lower right to the upper left can
be interpreted as a sequence of dust temperature.

Now we theoretically quantify the observed colour--colour
relation by adopting the calculations in
\citet{hirashita07}, who treat the dust heating by an ISRF
and the dust cooling by thermal radiation to calculate the
temperature distribution function using the framework
developed by \citet{draine01}. The physical quantities that
explain well the dust emission properties in the solar
neighbourhood are
adopted: the ISRF SED by \citet{mathis83}, the grain size
distribution by \citet{mathis77}, and the heat capacity of
grain materials by \citet{draine01}. The {emission
coefficient} of dust (silicate and graphite) are taken
from \citet{draine84}
for $\lambda\leq 100~\micron$ and extrapolated by
assuming a functional form proposed by \citet{reach95}
(i.e.\ $\beta$ smoothly changes from 1 to 2 around
$\lambda\sim 200~\micron$). As shown by
\citet{hirashita07}, a slight change in
$\beta$ affects the colour--colour sequence significantly.
To check the consistency with other nearby galaxies
in the colour--colour diagram, adopting the same emission
coefficient as adopted in \citet{hirashita07} is crucial
here.
We vary the ISRF with the spectral shape fixed, and denote
the ISRF intensity relative to the solar neighbourhood value
as $\chi$. We
refer to \citet{hirashita07} for the details of the framework
and some basic results.

We show the FIR colour--colour relation for various ISRF
intensity $\chi$ in Fig.\ \ref{fig:clr}. We only show
the results for graphite, since silicate follows the almost
identical FIR colour--colour relation to graphite
\citep{hirashita07}. We observe that the FIR colours
obtained for the individual grids are roughly explained
with $\chi =1$--30, although most of the points are located
systematically above the theoretical predictions on the
diagram. The lower values of $\chi$ correspond
to the lower-$T_\mathrm{LG}$ regions such as interarm
regions, and the higher values to the centre and the
bright spots in the spiral arms.

The above theoretical colour--colour sequence is correct
if the radiation field in a grid is approximated to be
uniform. As clearly seen in the H$\alpha$ image in
Fig.\ \ref{fig:image}, there are small-scale
star-forming regions, which should host warmer dust
because of high radiation field intensity. In order to
examine the effect of such a `contamination' of warm
dust, we show the
FIR colour--colour relation by mixing the results with
$\chi =1$ and $\chi =100$: the former value represents the
general ISRF, while the latter is taken as a representative
high radiation field value.
The fraction of the latter
(i.e.\ higher $\chi$) component is denoted as $f_\mathrm{h}$;
that is, the intensity is calculated by
\begin{eqnarray}
I_\nu =(1-f_\mathrm{h})I_\nu (\chi =1)+f_\mathrm{h}
I_\nu (\chi =100),
\end{eqnarray}
where $I_\nu (\chi )$ is $I_\nu$ as a function of $\chi$.
The result is shown in Fig.~\ref{fig:clr}. A slight
contamination of the higher-$\chi$ component with a fraction
of $f_\mathrm{h}=0.003$ significantly lift the colour
sequence upwards, explaining the upper part of the
FIR colour--colour relation of the individual grids.
This is because the 65 $\micron$ intensity responds
most sensitively to the higher $\chi$ component.
The shift of the FIR colour--colour relation by
the contamination of a higher $\chi$ component is consistent
with the conclusion by \citet{hibi06} and
\citet{hirashita07}. \citet{hibi06} called this upper
sequence `sub-correlation',
and the contamination effect `overlap effect'.

We also compare our results with the FIR colours of
the galaxies in the \textit{AKARI} FIS Bright Source
Catalogue in
Fig.\ \ref{fig:clr}. The analysis of these galaxies
has been done by \citet*{pollo10}. Since there are
an enormous number of galaxies, we only show the area
covered by the FIR Bright Source Catalogue galaxies. 
The redshifts of the sample are
small and do not affect the colours. We observe that the
FIR colour--colour relation in M\,81 is within the
consistent regime covered by the galaxy sample.
Note that the FIR colours of the FIS Bright Source
Catalog sample present
the global colours, not those in individual regions
within a galaxy. Thus, we confirm that the FIR colours of
individual regions within a galaxy is fundamental in
determining the global galaxy colours. The larger
scatter of the FIS Bright Source Catalogue sample may
be due to a
larger extent of the radiation field or a peculiarity
of dust emission properties in some galaxies.

In Fig.\ \ref{fig:clr}, we also show the colours of some
representative regions within the circles of 1-arcmin
radius as shown in Fig.\ \ref{fig:image}: the central
region (`C'), the spiral arms (`S1', `S2', and `S3'), and
the interarm regions (`I1' and `I2'). We calculate the
flux integrated for the circular regions, and take
the flux ratios to show the colours. The fluxes are listed
in Table \ref{tab:flux}. The central region
have the bluest colours, while the interarm regions
tend to have redder colours than the spiral arms: these
trends in colours are consistent with the temperature map
shown in Fig.~\ref{fig:dist_T_tau}.
Moreover, the central and interarm regions are located
{relatively} near to the
theoretical predictions with varying ISRFs 
(squares in Fig.~\ref{fig:clr}) on the colour--colour diagram, while
the spiral arms are shifted toward the theoretical
predictions with a mixture of general ISRF and a high
radiation field (asterisks in Fig.~\ref{fig:clr}). This
indicates that the ISRF
in the central and interarm regions are rather uniform
and occupied with general ISRFs with different intensities,
while there is a large variation in the radiation field in
the spiral arms
with a spatial scale well below 1 arcmin. It is
probable that small star-forming regions with strong
radiation fields
reside in spiral arms as seen in the H$\alpha$ image
(Fig.~\ref{fig:image}); this may be the reason for the
mixed colour features of the spiral arms.

\section{Discussion}\label{sec:discussion}

\subsection{FIR as a tracer of general ISRF}

The dust temperature derived from the global SED
(Section \ref{subsec:global}) is consistent with those
in the interarm regions and in the smooth regions
(excluding the warm spots) of the spiral arms.
This suggests that the global FIR SED is more
representative of the dust heated by the general ISRF
rather than by the
intense radiation in the star-forming regions.

That the FIR emission traces the general ISRF does
not necessarily mean that the FIR traces the smoothly
distributed old stellar population rather than the
newly formed stellar
population. It is generally hard to distinguish these
two components as pointed out by \citet{walterbos96}:
An elaborated treatment of multi-wavelength
star-formation indicators can be found in
\citet{hirashita03}, who find that $\sim 40$ per cent
of the FIR emission on average is associated with the
dust heated by stellar populations older than $10^8$ yr
in their nearby star-forming galaxy sample.

\subsection{Colour--colour diagram as a diagnostics
for the local heating source}\label{subsec:ccd}

The dust optical depth derived observationally depends
on the structures smaller than the spatial resolution.
If the ISRF is uniform in a grid used to derive the
optical depth, the dust
optical depth is reliable because of the
dust temperature estimated represents the real dust
temperature. However, if a grid is contaminated by
small-scale warm spots such as
a star-forming clouds, the temperature estimate is
biased to such warm spots. Because of this overestimate
of dust temperature, the dust optical
depth can be underestimated as pointed out
in Section~\ref{subsec:Tmap}. In other words, the FIR
emission is not a good tracer of dust mass
if the region focused on contains small-scale warm
regions contributing significantly to the FIR emission.

We have shown in Section \ref{subsec:clr} that the
contamination of warm regions can be identified by the
FIR colour--colour diagram. As we can clearly see in
Fig.~\ref{fig:clr}, the spiral arms (S1, S2, and S3),
which contain small-scale star-forming regions
(see the H$\alpha$ image in Fig.~\ref{fig:image}),
deviate from the the varying-$\chi$ sequence. The FIR
colours in the spiral arms are more consistent with
a mixture of a general ISRF with warm regions. Thus, we
suggest that the relatively small dust optical depth in
high-temperature regions of the
spiral arms is due to the overestimate of dust temperature
due to the contamination of warm spots possibly
associated with star-forming regions
(Section \ref{subsec:Tmap}).

In the central region marked by `C', the FIR colours
are close to the point of uniform radiation field with
$\chi\sim 10$. Thus, the region may be roughly
approximated with a single dust temperature. If this
is true, the optical depth in the central region
may reflect the real optical depth;
that is, the deficiency of dust optical depth
in the central
region with a radius of 1 arcmin (1.1 kpc)
may be real.
The uniformity of the radiation field
also implies that the heating by AGN is not a
dominant heating source in the central 1-kpc
region at least for the dust which can be traced in
the FIR bands. The H \textsc{i} gas is also
deficient in the central region \citep{rots75,braun95}.
Since the star formation activity in
the central region is strong as seen in the H$\alpha$
image, the radiation pressure or the thermal pressure
may have pushed the gas and dust outwards.
If we adopt 10 km s$^{-1}$
($\sim $ the sound velocity in the ionized regions) for
the typical velocity for the gas, a structure with a
size of 1.1 kpc can be formed on a time-scale of $10^8$
yr. Or some wave motion may be responsible for such a
structure \citep{rots75,lowe94}.

\subsection{Importance of \textit{AKARI} FIS bands
toward \textit{Herschel}}

The FIR wavelengths observed by the \textit{AKARI} FIS
bands (50--180\,$\micron$) cover the intensity peak of
the FIR SED. This means that the intensities
in the \textit{AKARI} FIS bands depends both on
the wavelength and the dust temperature in a strongly
nonlinear way. This strong nonlinearity is important
in the behaviour in the colour--colour diagram: not
only the strong dependence on $\chi$ but also
the large separation of the two sequences (i.e.\
varying $\chi$ and varying $f_\mathrm{h}$) is due to
such a nonlinearity.
{Moreover, the three (in fact four)
bands available for \textit{AKARI} FIS have advantage
in the colour--colour analysis.}

\textit{Herschel} is also suitable for the aim of
mapping dust emission in galaxies. In spite of its higher
spatial resolution, the warm spots associated with
star-forming regions cannot be resolved at the
distance of M\,81. Thus, the only way to show the
existence of such warm spots in the FIR is to plot the
FIR colours (Section \ref{subsec:ccd}).
Recently, \citet{bendo10} have observed M\,81 with
\textit{Herschel} over a wide wavelength range of
70--500 $\micron$.
The peak of the SED is at
$\lambda\la 160\,\micron$ for $T_\mathrm{LG}\ga 20$ K.
Thus, the
\textit{Herschel} PACS 100\,$\micron$ and 160\,$\micron$
bands \citep{poglitsch10} are suitable for covering
the peak.
The fact that similar temperatures are derived
by both \textit{AKARI} and \textit{Herschel}
(29 K, 20 K, 17 K for the nucleus, arm, and
interarm, respectively; \citealt{bendo10}) indicates
that the FIR SED can be interpreted consistently
from 90 $\micron$ to 500 $\micron$ with a single
temperature.
{Note that even with the \textit{Herschel} higher
resolution data, \textit{AKARI} data still have an
advantage that an enormous number of the
\textit{AKARI} All-Sky Survey data are available for
a comparison at the same wavelengths
(Section \ref{subsec:clr}).}

\section{Conclusion}\label{sec:conclusion}

We have investigated the properties of FIR emission in
a spiral galaxy, M\,81, by utilizing the \textit{AKARI}
imaging data at 65, 90, and 140 $\micron$. Combining
the images in the two long-wavelength bands
(90 and 140\,$\micron$), we have derived the dust
temperature map. The dust temperature is $\sim 25$\,K
in the centre and becomes lower in the outer part.
The dust temperature derived from the global
90 and 140\,$\micron$ intensities is 
$T_\mathrm{LG}=18.6^{+1.9}_{-1.6}$ K,
which reflects the dust temperatures in the interarm regions
or the spiral arms excluding the bright knots,
rather than those in the centre or in the bright knots.
Thus, the global dust temperature is more representative of
the dust heated by the general ISRF.
We have also shown
the 140 $\micron$ emission is more radially
extended than the 90 $\micron$ emission, which is
consistent with the radial dust temperature gradient.
The dust optical depth traced by the FIR emission is
$A_V=0.5$--2. If these values are divided by 2 with an
assumption that the stars are in the mid-plane on average,
the expected extinction for the stellar light is
$A_V=0.25$--1, consistent with the
the extinction derived from the
FIR-to-UV ratio ($A_V=0.57$).

We have also demonstrated that the FIR colour--colour
diagram is a useful tool to distinguish whether or not
individual regions within a galaxy is dominated by a
smooth ISRF or contaminated by warm spots associated with
star-forming regions. Based on this `tool', we conclude
that the bright regions in the spiral arms contain
small-scale warm regions possibly hosting the
H \textsc{ii} regions seen in the H$\alpha$ map.
This contamination of warm regions causes an
underestimate of dust optical depth (or dust column density).

Since even \textit{Herschel} cannot resolve individual
star-forming regions, the FIR colour--colour diagram
continues to be a useful tool to see if the dust
is predominantly heated by a general ISRF or the dust
heating by individual star-forming regions affects the
FIR emission.

\section*{Acknowledgments}
We are grateful to the anonymous referee for useful
comments and all members of the \textit{AKARI} project
for their continuous help and support. We thank
A. Pollo, P. Rybka, and T. T. Takeuchi for providing us
with the data of the \textit{AKARI} FIS Bright
Source Catalogue,
and T. Suzuki, B. T. Draine and H. Kaneda for helpful
discussions on the analysis and interpretation
of the data.
This research has made use of the NED, which is operated
by the Jet Propulsion Laboratory, California Institute
of Technology, under contract with the National Aeronautics
and Space Administration.
H.H. is supported by NSC grant 99-2112-M-001-006-MY3.


\bsp

\label{lastpage}


\begin{thebibliography}{}
\bibitem[\protect\citeauthoryear{Aannestad \& Kenyon}{1979}]{aannestad79}
    Aannestad, P. A., \& Kenyon, S. J. 1979, ApJ, 230, 771
\bibitem[\protect\citeauthoryear{Baggett, Baggett, \& Anderson}{1998}]{baggett98}
    Baggett, W. E., Baggett, S. M., \& Anderson, K. S. J. 1998,
    ApJ, 116, 1626
\bibitem[\protect\citeauthoryear{Bendo et al.}{2010}]{bendo10}
    Bendo, G., et al.\ 2010, A\&A, 518, L65
\bibitem[\protect\citeauthoryear{Braun}{1995}]{braun95}
    Braun, R. 1995, A\&AS, 114, 409
\bibitem[\protect\citeauthoryear{Buat et al.}{2005}]{buat05}
    Buat, V., et al.\ ApJ, 619, L51
\bibitem[\protect\citeauthoryear{Cheng et al.}{1997}]{cheng97}
    Cheng, K.P., Collins, N., Angione, R., Talbert, F., Hintzen, P.,
    Smith, E. P., Stecher, T., \& The UIT Team, Uv/visible Sky Gallery
    on CDROM
\bibitem[\protect\citeauthoryear{Dale et al.}{2001}]{dale01}
    Dale, D. A., Helou, G., Contursi, A., Silbermann, N. A., \& Kolhatkar, S.
    2001, ApJ, 549, 215
\bibitem[\protect\citeauthoryear{Dale et al.}{2007}]{dale07}
    Dale, D. A., et al.\ 2007, ApJ, 655, 863
\bibitem[\protect\citeauthoryear{Draine \& Anderson}{1985}]{draine85}
    Draine, B. T., \& Anderson, N. 1985, ApJ, 292, 494
\bibitem[\protect\citeauthoryear{Draine \& Lee}{1984}]{draine84}
    Draine, B. T., \& Lee, H. M. 1984, ApJ, 285, 89
\bibitem[\protect\citeauthoryear{Draine \& Li}{2001}]{draine01}
    Draine, B. T., \& Li, A. 2001, ApJ, 551, 807
\bibitem[\protect\citeauthoryear{Freedman et al.}{1994}]{freedman94}
    Freedman, W. L., et al.\ 1994, ApJ, 427, 628
\bibitem[\protect\citeauthoryear{Harper \& Low}{1973}]{harper73}
    Harper, D. A., Jr., \& Low, F. J. 1973, ApJ, 182, L89
\bibitem[\protect\citeauthoryear{Hibi et al.}{2006}]{hibi06}
    Hibi, Y., Shibai, H., Kawada, M., Ootsubo, T., \& Hirashita, H.
    2006, PASJ, 58, 509
\bibitem[\protect\citeauthoryear{Hirashita, Buat, \& Inoue}{2003}]{hirashita03}
    Hirashita, H., Buat, V., \& Inoue, A. K. 2003, A\&A, 410, 83
\bibitem[\protect\citeauthoryear{Hirashita, Hibi, \& Shibai}{Hirashita et al.}{2007}]{hirashita07}
    Hirashita, H., Hibi, Y., \& Shibai, H. 2007, MNRAS, 379, 974
\bibitem[\protect\citeauthoryear{Hirashita et al.}{2008}]{hirashita08}
    Hirashita, H., Kaneda, H., Onaka, T., \& Suzuki, T. 2008, PASJ, 60,
    S477
\bibitem[\protect\citeauthoryear{Inoue, Hirashita, \& Kamaya}{2000}]{inoue00}
    Inoue, A. K., Hirashita, H., \& Kamaya, H. 2000, PASJ, 52, 539
\bibitem[\protect\citeauthoryear{Kawada et al.}{2007}]{kawada07}
    Kawada, M., et al.\ 2007, PASJ,,  59, S389
\bibitem[\protect\citeauthoryear{Kennicutt}{1998}]{kennicutt98}
    Kennicutt, R. C., Jr.\ 1998, ARA\&A, 36, 189
\bibitem[\protect\citeauthoryear{Li \& Draine}{2001}]{li01}
    Li, A., \& Draine, B. T. 2001, ApJ, 554, 778
\bibitem[\protect\citeauthoryear{Lowe et al.}{1994}]{lowe94}
    Lowe, S. A., Roberts, W. W., Yang, J., Bertin, G., \&
    Lin, C. C. 1994, ApJ, 427, 184
\bibitem[\protect\citeauthoryear{Mathis, Rumpl, \& Nordsieck}{1977}]{mathis77}
    Mathis, J. S., Rumpl, W., \& Nordsieck, K. H. 1977, ApJ, 217, 425
\bibitem[\protect\citeauthoryear{Mathis, Mezger, \& Panagia}{1983}]{mathis83}
      Mathis, J. S., Mezger, P. G., \& Panagia, N. 1983, A\&A, 128, 212
\bibitem[\protect\citeauthoryear{Murakami et al.}{2007}]{murakami07}
    Murakami, H., et al.\ 2007, PASJ, 59, S369
\bibitem[\protect\citeauthoryear{Nagata et al.}{2002}]{nagata02}
    Nagata, H., Shibai, H., Takeuchi, T. T., \& Onaka, T. 2002, PASJ,
    54, 695
\bibitem[\protect\citeauthoryear{P\'{e}rez-Gonz\'{a}lez et al.}{2006}]{perez06}
    P\'{e}rez-Gonz\'{a}lez, P. G., et al.\ 2006, ApJ, 648, 987
\bibitem[\protect\citeauthoryear{Pilbratt et al.}{2010}]{pilbratt10}
    Pilbratt, G. L., et al.\ 2010, A\&A, 518, L1
\bibitem[\protect\citeauthoryear{Poglitsch et al.}{2010}]{poglitsch10}
    Poglitsch, A., et al.\ 2010, A\&A, 518, L2
\bibitem[\protect\citeauthoryear{Pollo, Rybka, \& Takeuchi}{Pollo et al.}{2010}]{pollo10}
    Pollo, A., Rybka, P., \& Takeuchi, T. T. 2010, A\&A, 514, A3
\bibitem[\protect\citeauthoryear{Reach et al.}{1995}]{reach95}
    Reach, W. T. et al., 1995, ApJ, 451, 188
\bibitem[\protect\citeauthoryear{Rice et al.}{1988}]{rice88}
    Rice, W., Lonsdale, C. J., Soifer, B. T., Neugebauer, G.,
    Kopan, E. K., Lloyd, L., de Jong, T., \& Habing, H. J. 1988,
    ApJS, 68, 91
\bibitem[\protect\citeauthoryear{Rots}{1975}]{rots75}
    Rots, A. H. 1975, A\&A, 45, 43
\bibitem[\protect\citeauthoryear{Sauvage et al.}{2010}]{sauvage10}
    Sauvage, M., et al.\ 2010, A\&A, 518, L64
\bibitem[\protect\citeauthoryear{Shibai, Okumura, \& Onaka}{1999}]{shibai99}
    Shibai, H., Okumura, K., \& Onaka, T. 1999, in Nakamoto T. ed.,
    Star Formation 1999, Nobeyama Radio Observatory, Nobeyama, p.\ 67
\bibitem[\protect\citeauthoryear{Shirahata et al.}{2008}]{shirahata08}
    Shirahata, M., et al.\ 2008, PASJ, 61, 737
\bibitem[\protect\citeauthoryear{Suzuki et al.}{2007}]{suzuki07}
    Suzuki, T., Kaneda, H., Nakagawa, T., Makiuti, S., \& Okada, Y.
    2007, PASJ, 59, S473
\bibitem[\protect\citeauthoryear{Suzuki et al.}{2010}]{suzuki10}
    Suzuki, T., Kaneda, H., Onaka, T., Nakagawa, T., \& Shibai, H. 2010,
    A\&A, in press (arXiv: 1006.2018)
\bibitem[\protect\citeauthoryear{Takeuchi et al.}{2010}]{takeuchi10}
    Takeuchi, T. T., Buat, V., Heinis, S., Giovannoli, E.,
    Yuan, F.-T., Iglesias-P\'{a}ramo, J., Murata, K. L., Burgarella, D.
    2010, A\&A, 514, A4
\bibitem[\protect\citeauthoryear{Verdugo, Yamamura, \& Pearson}{Verdugo et al.}{2007}]{verdugo07}
    Verdugo, E., Yamamura, I., \& Pearson C. P., 2007, AKARI FIS Data
    User Manual Version 1.3 (http://www.ir.isas.ac.jp/ASTRO-F/Observation/)
\bibitem[\protect\citeauthoryear{Walterbos \& Greenawalt}{1996}]{walterbos96}
    Walterbos, R. A. M., \& Greenawalt, B. 1996, ApJ, 460, 696
\bibitem[\protect\citeauthoryear{Weingartner \& Draine}{2001}]{weingartner01}
    Weingartner, J. C., \& Draine, B. T. 2001, ApJ, 548, 296
\bibitem[\protect\citeauthoryear{Yamamura et al.}{2009}]{yamamura10}
    Yamamura, I., et al.\ 2009, in Onaka T., White G., Nakagawa T.,
    Yamamura I. ed., ASP Conf.\ Ser.\ Vol.\ 418, AKARI, a
    Light to Illuminate the Misty Universe. Astron.\ Soc.\ Pac.,
    San Francisco, p.\ 3
\end{thebibliography}
\end{document}